\documentclass[onecolumn]{article}
\usepackage[noadjust]{cite}
\usepackage{hyphenat}
\usepackage{siunitx}
\usepackage{chemformula}
\usepackage[T1]{fontenc}
\usepackage{hyperref}                                   %hyperlink your equations and figures and sorts

\title{Unidirectional mode selection in bistable quantum cascade ring lasers}

\author{Sara Kacmoli\textsuperscript{1,*}, Deborah L. Sivco\textsuperscript{2}, Claire F. Gmachl\textsuperscript{1}}
\date{
	$^1$Department of Electrical and Computer Engineering, Princeton University, Princeton, NJ 08544 USA\\ %\texttt{\{auth1, auth3\}@org1.edu}\\%
	$^2$Current address: Trumpf Photonics, Inc., 2601 US Highway 130, Cranbury, NJ 08512 USA\\ $^*$\texttt{skacmoli@princeton.edu}\\[2ex]%
}
\begin{document}

\maketitle

\begin{abstract}
Ideal ring resonators are characterized by travelling-wave counterpropagating modes, but in practice travelling waves can only be realized under unidirectional operation, which has proved elusive. Here, we have designed and fabricated a monolithic quantum cascade ring laser coupled to an active waveguide that allows for robust, deterministic and controllable unidirectional operation. Spontaneous emission injection through the active waveguide enables dynamical switching between the clockwise and counterclockwise states of the ring laser with as little as 1.6$\%$ modulation of the electrical input. We show that this behavior stems from a perturbation in the bistable dynamics of the ring laser. 
In addition to switching and bistability, our novel coupler design for quantum cascade ring lasers offers an efficient mechanism for outcoupling and light detection.
\end{abstract}

%\setboolean{displaycopyright}{true}

%%%%%%%%%%%%%%%%%%%%%%%%%%  body  %%%%%%%%%%%%%%%%%%%%%%%%%%

\section{Introduction}

Ring lasers have been and continue to be investigated for their rich dynamics in a variety of geometries and material platforms. Numerous phenomena based on ring lasers have been numerically studied and experimentally observed, such as unidirectional and bidirectional operation, alternate oscillations, bistability and chaos~\cite{Sorel_unibistability, Sorel_opregimes, Perez_SRLdynamics, Sande_SRLdynamics, Sunada_ASEbistability, Nguimdo_chaos}. These phenomena result from the nonlinear interaction of the clockwise (CW) and counterclockwise (CCW) fields. These two fields, which in an ideal resonator do not couple to each other, give rise to the travelling-wave nature of the total field in a ring laser. This is in contrast to Fabry-Perot (FP) lasers, where the electric field exhibits a standing-wave pattern leading to spatial hole burning (SHB) of the gain medium. In lasers where the gain recovery time is fast, such as quantum cascade lasers (QCLs), SHB effects are dominant and lead to multimode instabilities~\cite{Gordon_SHB}. Although ring resonators seem to evade SHB, pure travelling waves are only possible in an ideal cavity; in a realistic cavity, even small imperfections will perturb the rotational invariance of the counter-propagating modes and pin them to a standing-wave pattern similar to FP lasers. Thus, the only realistic way to obtain travelling waves in a ring laser is by ensuring unidirectional operation where only one of the two modes, CW or CCW, is allowed to lase at a time. Unidirectional operation of ring QCLs has been shown to be necessary for the demonstration of active mode locking and mid-infrared, ring-based frequency combs~\cite{malara_extRQCL, Meng_RQCLFC}. Previous work on QCLs aimed directly at mode selection between CW and CCW modes has relied on an "S-shaped" waveguide design to favor one mode~\cite{Nshii_uniRQCL} or external cavities where the gain medium is placed within a free-space-coupled ring cavity~\cite{malara_extRQCL, Revin_activeml}.

In this work, we show robust and deterministic unidirectional operation from a monolithic racetrack QCL. Instead of a specialized waveguide design of the ring laser itself, which permanently favors one mode, we rely on an active waveguide evanescently coupled to the racetrack to perform mode selection and switching, which allows for controllable operation in \textit{both} directions. This in turn, may enable denser integration and greater functionality of mid-infrared photonic integrated circuits. 

Switching between the two modes of a semiconductor ring laser has also been used to demonstrate random number generation~\cite{Sunada_ASEbistability}. The fast carrier lifetime of QCLs ($\sim$ ps) may be used in such applications to improve the speed of this operation. In addition, we show that our ring laser exhibits a region of bistability, a dynamical regime that has not yet been explored in mid-infrared QCLs. The bistable nature of the ring allows for full switching between the two modes with small input perturbations. 

Our QCL operates at  $\lambda \approx$ \SI{8}{\um}, a wavelength where the fabrication of an evanescent coupler becomes challenging because the active region is buried deep in the QC layers thus requiring a complex etching process. Unlike previous work, which focuses on short wavelengths ($\lambda <$ \SI{5}{\um}), \cite{Nshii_uniRQCL, jung_integration} we fabricate an evanescent coupler that does not rely on any specialized growth or processing techniques. 

Lastly, ring QCLs have recently been studied as emerging platforms for frequency combs~\cite{Meng_RQCLFC, Piccardo_phaseturb, Meng_soliton}. In these studies, light extraction is done by sensing the weak scattering off the waveguides due to bending losses. While this technique aids in minimizing backscattering between the two counter-propagating modes, it may pose a limitation on the practical use of these devices. In addition to controllable mode selection and switching, the active waveguide in our system also serves as an efficient outcoupling mechanism for the racetrack laser.

\section{Design and Fabrication}

We have fabricated a racetrack resonator coupled to a bus waveguide on a conventional $\lambda \approx$ \SI{8.8}{\um} QCL wafer based on a double-phonon resonance design \cite{Liu_qcl2phonon}. The \textit{active} bus waveguide differs from the more commonly utilized \textit{passive} bus waveguide in two key ways. First, it has intersubband transitions at the same wavelength that the racetrack emits; thus, to overcome absorption, the waveguide must be pumped to transparency. Second, due to the same transitions and depending on the pumping level, it has its own spontaneous emission which is coupled into the racetrack and is responsible for the switching properties we will discuss here. 

\begin{figure}[!ht]
\centering\includegraphics[width=8.7cm]{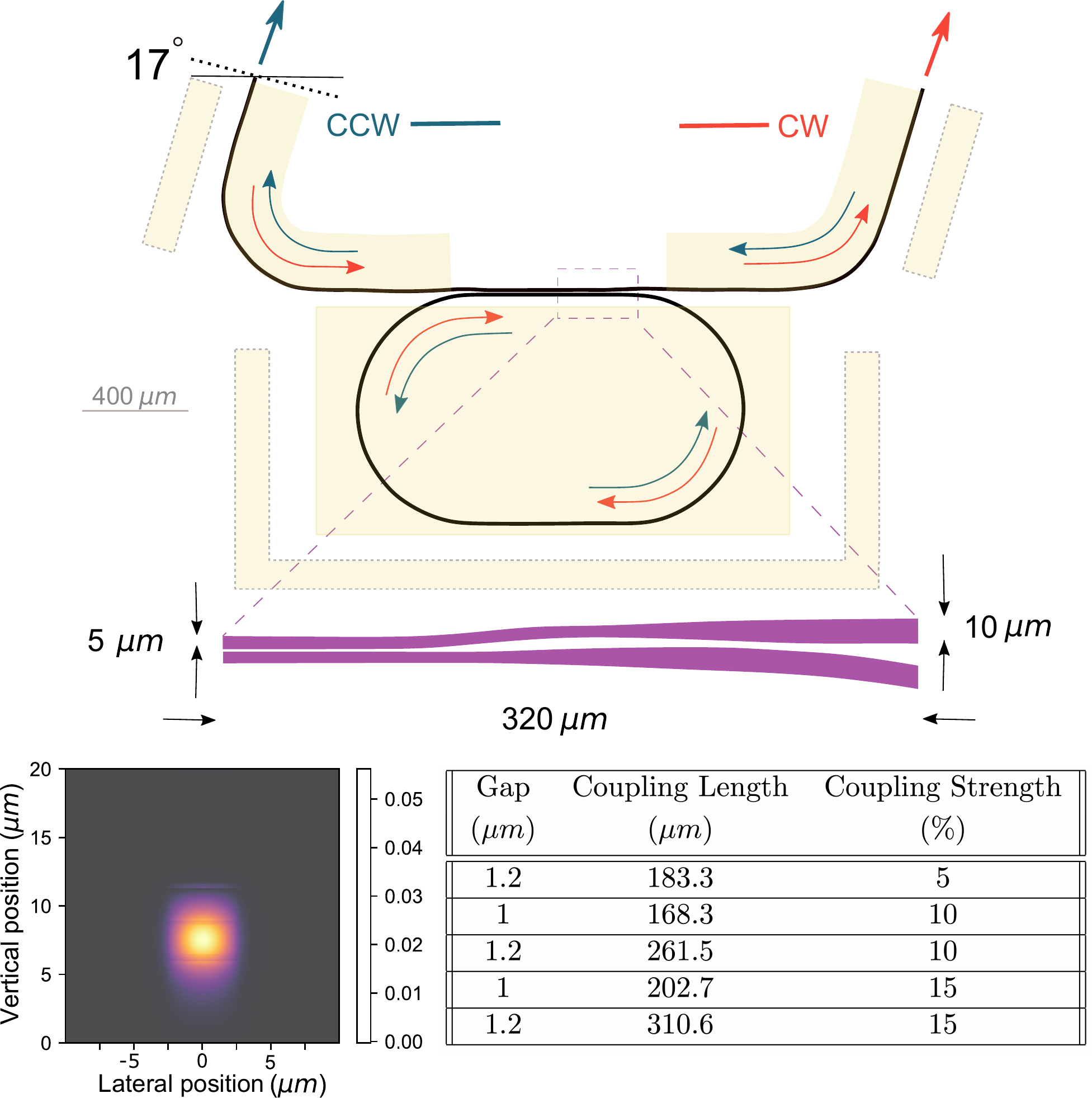}
\caption{Device schematic showing the racetrack laser ridge and the evanescently-coupled waveguide with facets tilted at $17^{\circ}$. The dotted purple box shows a magnified view of the taper and s-bend employed on the ridges in order to enhance their coupling. The blue and red arrows illustrate how light is coupled in and out of the racetrack. Yellow regions represent a schematic of the top and ground metal contact pads. The bottom left panel shows the calculated mode intensity of a \SI{5}{\um} ridge waveguide. The bottom right panel offers a few combinations of gap width and coupling length leading to particular coupling strengths as calculated by CMT. The devices discussed here correspond to the last set of listed parameters.}
\label{coupler}
\end{figure}

The schematic of our system is shown in Fig.~\ref{coupler}. The two arms of the waveguide have separate metal contacts and are thus controlled independently. The coupling mechanism in this system is evanescent coupling where a small gap is deep etched between the ridges. To determine the size of the gap as well as the length of the coupler we use coupled mode theory (CMT) simulations based on the computed fundamental mode. Our CMT simulations show that significant coupling can only occur for relatively thin ridges. In our case we choose a \SI{5}{\um} ridge width because the mode is less confined laterally thereby expanding outside the waveguide core and enhancing evanescent coupling. For ease of fabrication in the following processing steps, we adiabatically taper our waveguide width to \SI{10}{\um} outside of the coupling region. This also helps to limit loss due to weak confinement and proximity of the mode to the lossy metal on the sidewalls. The detailed design of the coupler is shown in Fig.~\ref{coupler}. We have considered several combinations of gap and coupling length and their corresponding coupling strength is given in the table in Fig.~\ref{coupler}.

\begin{figure}[!ht]
\centering\includegraphics[width=8.8cm]{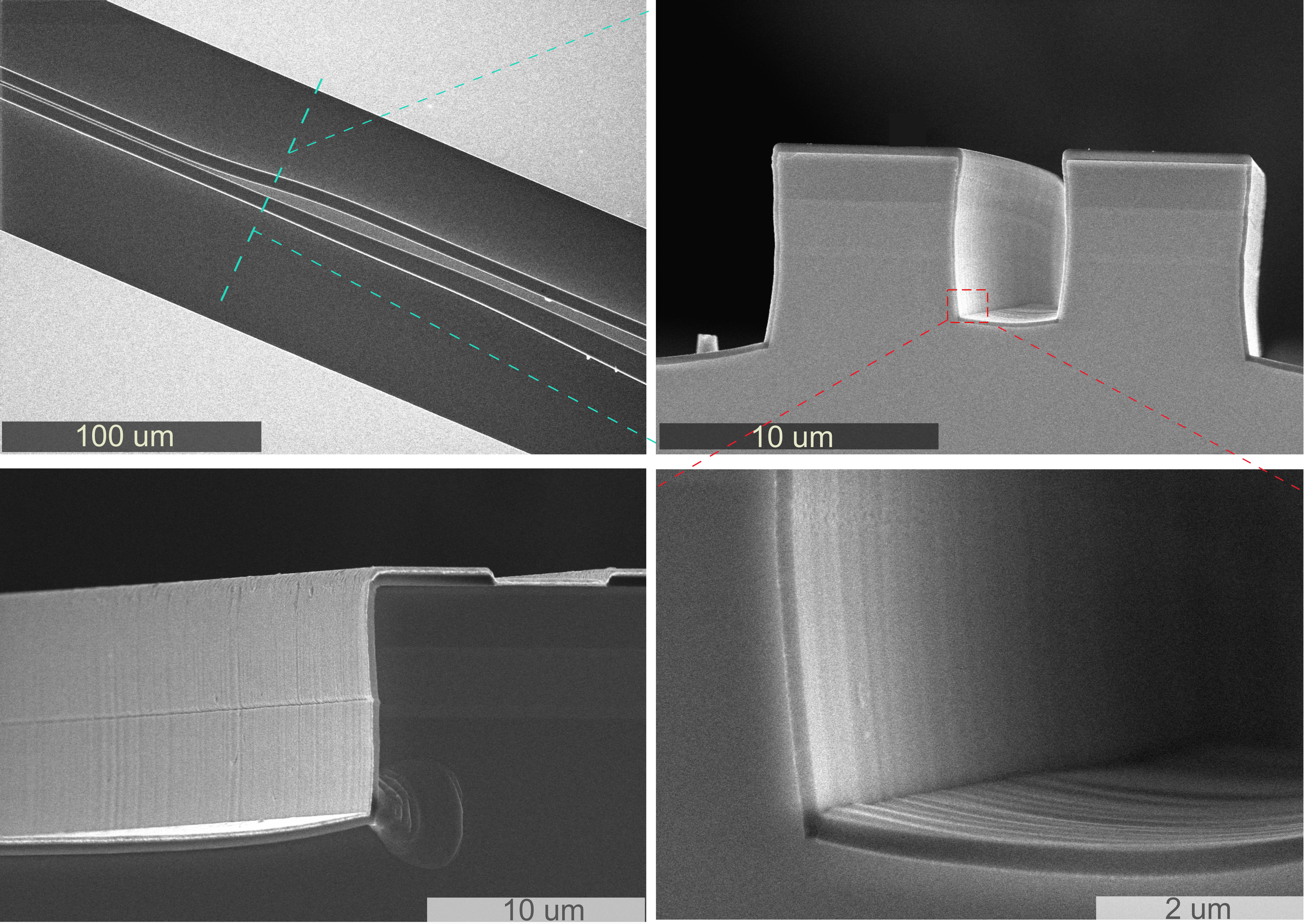}
\caption{Scanning electron microscope (SEM) images of the device, focusing on the coupler. Top left: A top-view of the coupling region and taper. Top right: Cross-sectional view along the taper. The darker contour along the ridges is the insulating \ch{SiNx} layer. The coupling region is kept free of metal. Bottom left: Cross-sectional view of the ridge waveguide elsewhere in the device with a dielectric window and \ch{Ti}/\ch{Au} metal layer for electrical contact. Bottom right: Magnified view showing the etch profile and minimal, highly sub-wavelength roughness on the waveguide sidewalls.}
\label{fab}
\end{figure}

The chosen physical dimensions are compatible with conventional lithography techniques. The most challenging step in the fabrication is deep etching of the gap with an aspect ratio $\geq 5$ while still maintaining vertical sidewalls and minimal roughness. We utilize a multistep inductively coupled plasma – reactive ion etching (ICP-RIE) process based on \ch{Cl2} and \ch{BCl3}. The ICP-RIE process consists of many parameters, which we optimize for minimal sidewall roughness. Fig.~\ref{fab} shows scanning electron microscope (SEM) images of various elements of the device to especially showcase the etch characteristics. To reduce bending losses, all bend radii are designed to be larger than \SI{450}{\um}. To avoid feedback from reflection, the waveguide arms are angled $17^{\circ}$ with respect to the cleaved facet which lowers their reflectivity to $\approx 0.01$ \cite{Ahn_reflectivity, Aung_reflection}. We route the waveguides to the same side of the chip for ease of measurement, while ensuring both arms have practically the same path length to match their optical properties. Finally, an important design consideration when working with a system of two or more elements on the same wafer is electrical crosstalk due to a non-negligible ground impedance. To mitigate this issue, instead of the more conventional common, substrate-side ground used in QCLs, we use separate top ground pads for the racetrack laser and each waveguide arm.

\section{Temporal Dynamics and Mode Switching}
The racetrack laser supports two counter-propagating modes which in an ideal cavity are best described by travelling waves. Although we optimize the fabrication process, even minimal sidewall roughness will lead to backscattering and coupling between the modes. One way to preserve travelling waves in the cavity is if only one of the counter-propagating modes is allowed to lase at a time. A mechanism which allows us to select the dominant mode accurately and to efficiently switch between them is a useful tool to achieving controllable unidirectional lasing in the racetrack. 
To distinguish the two modes experimentally, in our measurements we use a thermoelectrically (TE)-cooled HgCdTe (MCT) detector with a 1x\SI{1}{\mm} active area along with a pair of \ch{ZnSe} lenses of 1.5" focal length. By translating the detector across the length of the device, we are able to fully resolve the two peaks emitted from both waveguide arms, which correspond to the CW and CCW modes. All measurements are performed in pulsed mode using \SI{300}{\ns} pulses with \SI{50}{\kHz} repetition rate. The device is thermally stabilized at room temperature with a TE cooler.

\begin{figure}[ht!]
\centering\includegraphics[width=9cm]{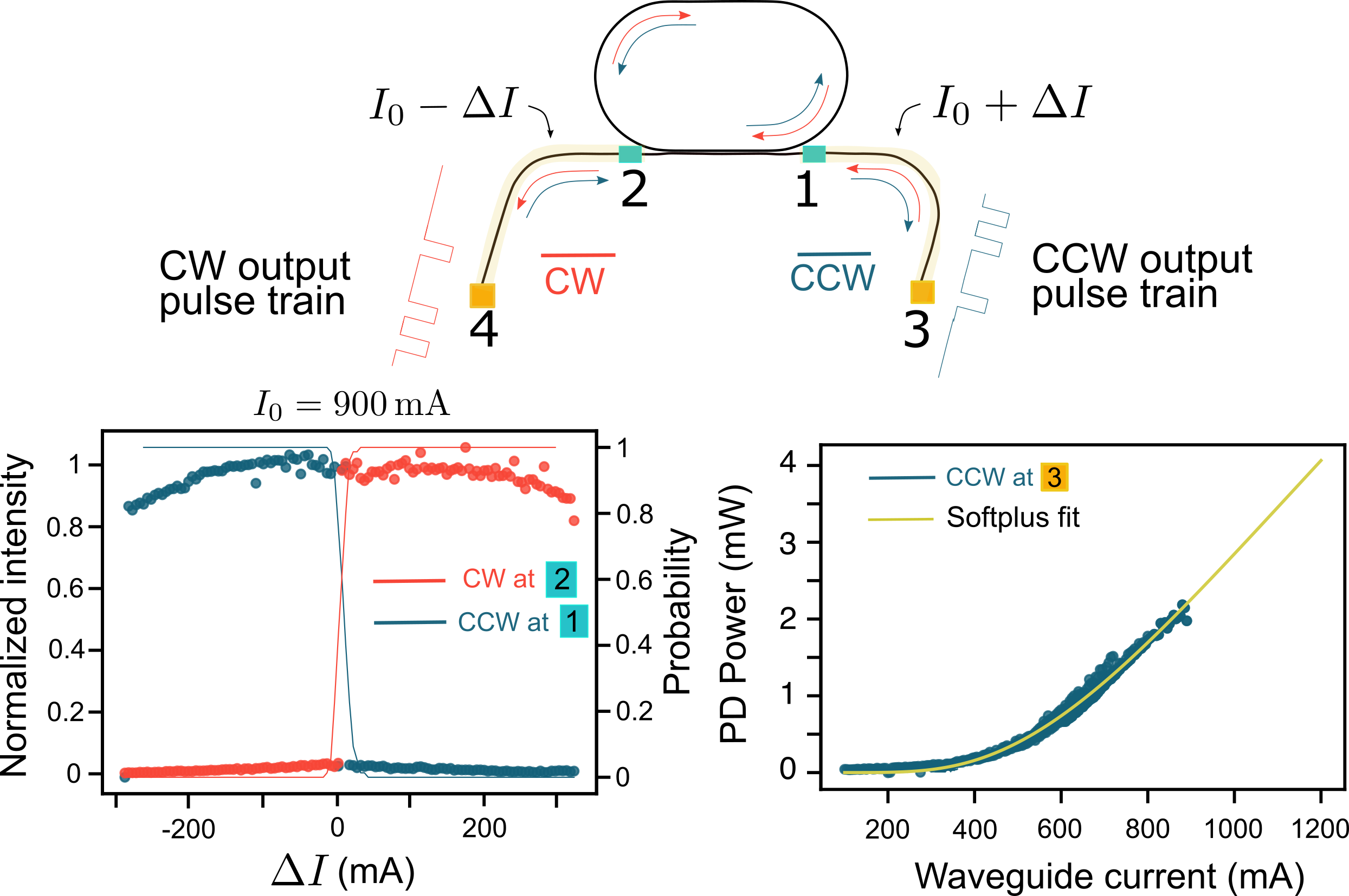}
\caption{Top: System schematic. Bottom left: CW and CCW normalized intensity given as a function of injected current imbalance between the two waveguide arms as illustrated in the system schematic. The CW (CCW) mode shown in dotted red (blue) dominates for positive (negative) $\Delta I$. This is the signal right as it exits the racetrack, represented by the turquoise markers at ports 1 and 2 in the schematic. In solid red (blue) we show the measured probability that the mode is CW (CCW) from 15 consecutive measurements. Bottom right: CCW power measured by an MCT photodetector as it exits the waveguide (noted by an orange marker at ports 3) along with a Softplus fit.}
\label{switching}
\end{figure}

We are able to detect the signal from the racetrack laser exiting the waveguide facets at ports 3 and 4 (shown in the top schematic of Fig.~\ref{switching}) even without any pump applied to the waveguide arms, although the signal is attenuated due to significant absorption along each arm. The racetrack threshold current is $\approx$ \SI{1.25}{\A} or \SI{2.89}{\kA/{\cm}^2}. Our measurements show that the racetrack laser is unidirectional; however, at each current pulse, it selects between the CW and CCW mode randomly. Under pulsed measurements, the output thus fluctuates between the two modes.

We show that a small amount of amplified spontaneous emission injected into one of the two arms favors one of the modes. In our experimental demonstration of mode selection we differentially pump the two waveguide arms taking advantage of the fact that spontaneous light emitted by each arm only couples to one of the two possible modes. Similarly, each mode will couple to a different arm of the active waveguide which allows us to distinguish them with our MCT detector. The schematic in Fig.~\ref{switching} illustrates these mode interactions with color-coded arrows. In our experiment we pump the two waveguide arms differentially by applying a variable current $I=I_{0} \pm \Delta I$ to each arm. This way the average applied current, $I_0$, at any point during our sweep remains constant. With the external MCT detector we then capture the signal at ports 3 and 4 (marked in orange in the schematic). The racetrack signal exiting the facets at 3 and 4 has interacted with the waveguide arms which provide pump-dependent absorption or amplification. An experimental measurement of this interaction is shown on the bottom right panel of Fig.~\ref{switching} which depicts the CCW output at port 3 as a function of the current applied on the left waveguide arm in the same schematic (which favors CCW operation). For up to $\approx$ 400 mA of applied current, absorption is the dominant effect; however, beyond this point, gain in the waveguide takes over and the output is amplified with increasing current. Measuring this characteristic gain curve allows us to calculate the output right as it exits the racetrack (schematically depicted by the turquoise ports 1 and 2). We use a Softplus function to fit the gain curve and then use this fit to calculate what the measured signal exiting ports 3 and 4 looks like at ports 1 and 2. This allows us to investigate and display the mode selection mechanism more clearly without the additional effects of absorption or amplification in the waveguide.

We have shown on the bottom left panel of Fig.~\ref{switching} the measured CW and CCW signals corrected using the fit discussed above as a function of the current differential between the two waveguide arms. This is the output immediately after it is outcoupled from the racetrack \textit{before} it travels through the length of the waveguide arms.
The lasing mode (CW or CCW) is only dependent on which arm has a higher pump. Following notation in the schematic of Fig.~\ref{switching}, for $\Delta I > 0$, the right arm has a higher spontaneous emission, favoring the CW mode. Conversely, for $\Delta I < 0$, the left arm's stronger emission favors the CCW mode.
For a small region around $\Delta I = 0$, the mode is chosen at random.
% If more spontaneous emission is coupled in to CW (CCW) mode by applying current to the corresponding arm, the CW (CCW) mode will dominate.
In the same plot we show the probability that the mode is CW (red, solid) or CCW (blue, solid) as computed from a representative set of 15 consecutive measurements for any given applied current. The intensity of the dominant mode remains relatively constant as we sweep $\Delta I$. 
Any light injected into the racetrack from the waveguide arms in these measurements is due to spontaneous emission. In fact, the two waveguide arms form a FP cavity and therefore \textit{can} achieve lasing if pumped together. However, in any of these measurements, the bus waveguide is well below lasing because we only pump one arm at a time which leads to significant loss in the bus cavity. Together with the $17^{\circ}$ facets, this loss leads to a high lasing threshold ($\approx$ \SI{7.84}{\kA/{\cm}^2} or \SI{3.36}{A} if both arms are pumped simultaneously). 

A mode selection mechanism based on spontaneous emission injection is consistent with our measurements. We have also considered the role that feedback of the racetrack signal from the facets back into the racetrack may play in our system; however, we have found this effect to be weak considering the combined effect of the weak reflection from the facets and a double pass through the $15\%$ coupler.

% \fxwarning{Starting with ``Of course''... until now, you can simply state that the you have measured the gain curve of the WG section. As current is pumped, you can amplify the output of the RT, a very desirable property in such systems...}

\begin{figure}[h!]
\centering\includegraphics[width=9cm]{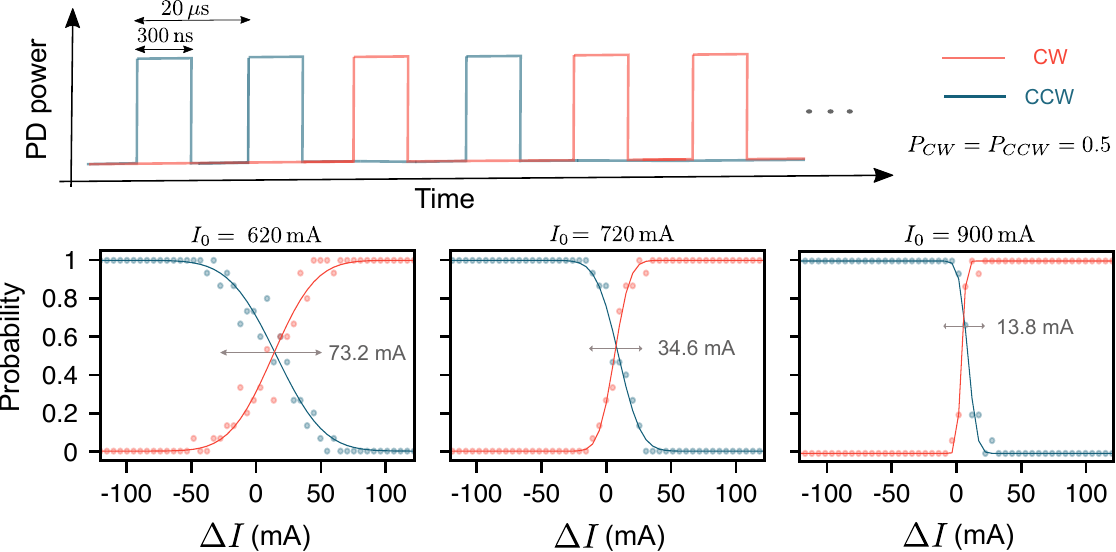}
\caption{Top: Schematic of a received pulse train used to measure the probability of each mode appearing. In this example, the probability $P_{CW} = P_{CCW} = 0.5$. Bottom: Probability of mode selection for the CW (red) and CCW (blue) modes. The solid line is an error function fit of the raw data (displayed as dots). The mean current increases from left to right from \SI{620}{\mA} to \SI{900}{\mA}. Required switching current, calculated as the width of the $10\%$-$90\%$ interval of each fit, decreases from \SI{73.2}{\mA} to \SI{13.8}{\mA} from left to right. The racetrack current in these measurements is held constant at \SI{1.81}{A} (\SI{4.19}{\kA/{\cm}^2}).}
\label{probability}
\end{figure}

Having shown that we are able to select one of the two modes based on the current differential $\Delta I$, we now examine the effectiveness of switching between modes. We measure the amount of current imbalance needed to switch for various average currents $I_0$ applied to the waveguide arms. To this end, we assess the probability for each mode to dominate by capturing a representative train of 15 consecutive measurements from the MCT detector as shown schematically in Fig.~\ref{probability}. We only observe two possible responses from the photodetector. For example, if we are observing at the CW port, we see a high photodetector response if the mode is CW or zero response if the mode is CCW and never in between. The probability $P_{CW(CCW)}$ is then defined as the number of pulses measured in CW (CCW) mode divided by the total number of pulses, in our case 15. The results of this analysis are shown in Fig.~\ref{probability} for three different values of $I_0$. Note the crossover happens around $\Delta I \approx \SI{0}{\mA}$. It is always slightly shifted to one side, which we explain based on a slight imbalance in the lengths of the two waveguide arms. Another key detail is the span of the bistable region (where probabilities of each mode are comparable). To determine this span we fit our data using the error function and calculate the width going from $10\%$ to $90\%$ probability. For $I_0$ = \SI{620}{\mA} this region spans $\approx$ \SI{73.2}{\mA}. As $I_0$ increases to \SI{900}{\mA}, the span of this region decreases to \SI{13.8}{\mA}. %\todo{explain this somehow}

\section{Bistability Mechanism}
\begin{figure}[h!]
\centering\includegraphics[width=9cm]{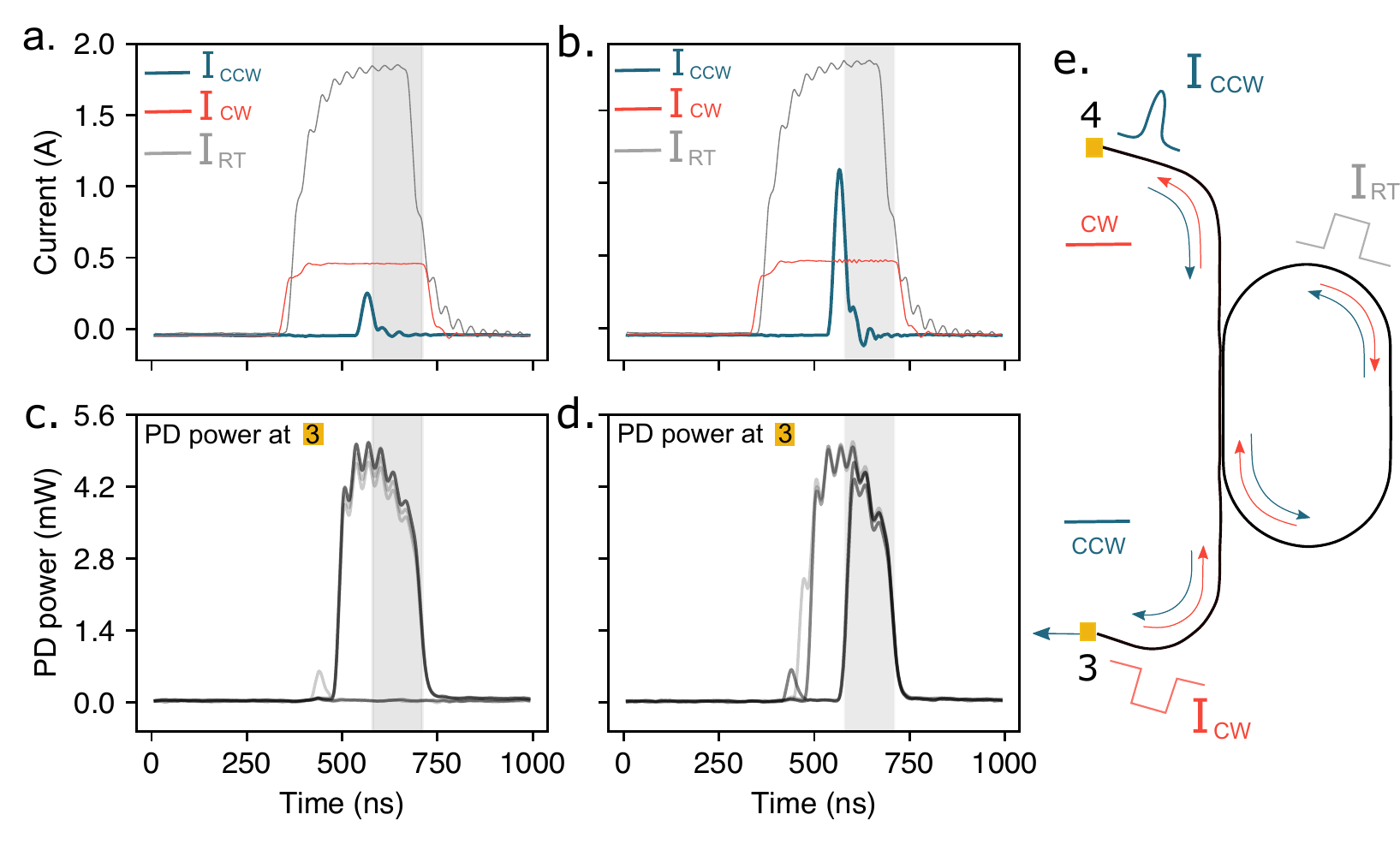}
\caption{Top: Input pulses into the racetrack and each waveguide arm. \textbf{(a)} The spike coupling to the CCW mode is smaller than the CW background.  \textbf{(b)} The spike coupling to the CCW mode is larger than the CW background. \textbf{(c)}, \textbf{(d)} 15 superimposed photodetector traces at port 3 corresponding respectively to pumping scenarios in (a), (b). Note the difference in the region shaded in gray between plots (c) and (d). When the short pulse exceeds pumping in the opposite arm, the CCW mode always dominates. \textbf{(e)} Device schematic along with mode interactions. 
}
\label{bistability}
\end{figure}

Thus far we have shown that we can select the direction in which the racetrack laser emits by modulating the current differential into the waveguide arms. However, a question arises on the mechanism that enables this phenomenon. Is it necessary to continuously pump a particular arm of the waveguide if we wish to maintain the directionality of the racetrack emission or is it sustained by the racetrack dynamics itself? To answer this question we conduct the experiment shown in Fig. \ref{bistability}.
The setup is described by the upper panels of the figure. A \SI{300}{\ns} pulse of $\approx$ \SI{1.81}{\A} is applied to the racetrack. At this current, the racetrack is 1.4 times above the lasing threshold. Midway through the racetrack pulse, we apply a short current pulse of width $\approx$ \SI{30}{\ns} and current up to \SI{1.15}{A} to the waveguide arm that favors CCW operation. In the opposite arm we apply a \SI{300}{\ns} pulse of $\approx$ \SI{480}{\mA} just above the necessary current input for absorption mitigation.
% in order to mitigate absorption. We verify that the presence of this pulse (depicted in red) does not change the behaviour of the racetrack, but merely compensates the absorption as we have described in the previous section.
Observing the photodetector response during each pulse before and after the short \SI{30}{\ns} pulse input is one way to address the question we have posed above.
On the lower left plot we show the output on the CCW side (port 3) measured on a TE-cooled MCT detector when the input CCW spike peak (in blue) is \textit{lower than} the longer CW pulse (in red). We plot 15 consecutive measurements which reveal a seemingly random selection of the lasing direction.
% The presence of a pulse signal measured from the detector indicates dominance of CCW mode; on the other hand, when there is no signal, the CW mode dominates.
When the laser's CCW mode dominates, we measure a pulsed signal on the detector; conversely, when the CW mode dominates, no signal is measured, as shown in Fig.~\ref{bistability}.
On the lower-right plot we show photodetector measurements when the spike injected into the arm favoring CCW operation is \textit{higher than} the CW background injected into the other arm.
% Precisely at the time of the input spike, the output of the racetrack becomes deterministic selecting the CCW mode every time.
We observe that this input perturbation causes the racetrack laser to switch to the CCW mode deterministically, regardless of the initial lasing mode.
% Note how there is no 'zero' response from the detector in the second half of the pulse which shows there is no CW mode, only CCW in all 15 measurements.
This can be seen for the section of the pulse highlighted in gray, where all of the measurements show a high signal on the detector, corresponding to the CCW mode.
Moreover, the laser remains fixed at the CCW mode, even after the perturbation that caused the switch dissipates.
This sheds some light on the mode selection mechanism.
% While the external waveguide arm provides a 'seed' for one particular racetrack mode to dominate, it does not need to be continuously pumped for this particular mode to continue living within the racetrack.
% Rather, it is the bistable nature of the racetrack dynamics itself that allows for unidirectional mode selection.
While pumping the external waveguide arm favors one particular racetrack mode to dominate, the selected mode is self-sustaining thereafter. This demonstrates the bistable nature of the racetrack dynamics which allows for unidirectional mode selection.

\section{Spectral Characteristics}

\begin{figure}[!ht]
\centering\includegraphics[width=9cm]{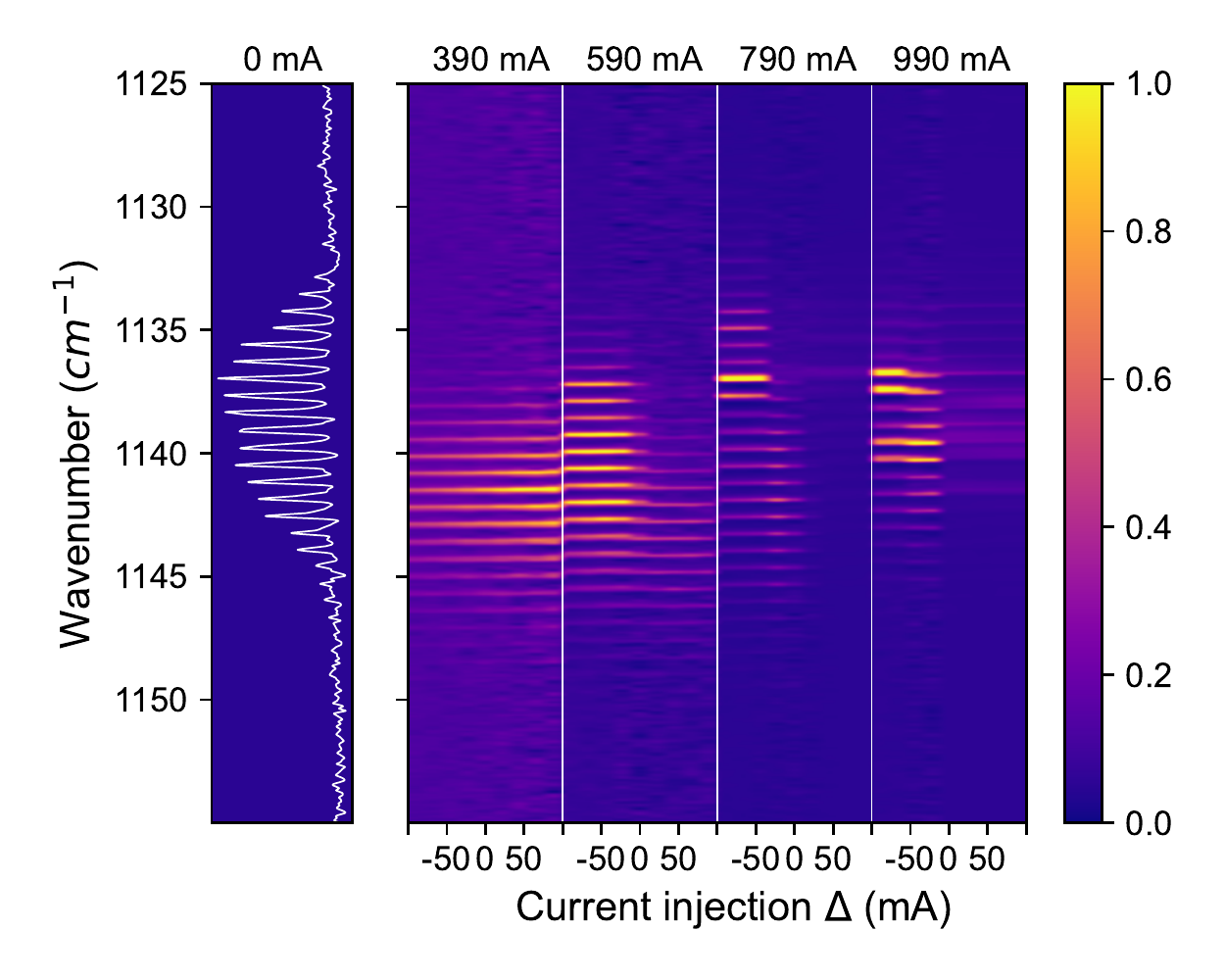}
\caption{Spectral characteristics of the device. The left panel corresponds to the spectrum of the racetrack when no current is applied to the waveguide arms. For the other panels, the average applied current $I_0$ is given at the top and the spectra are given as a function of $\Delta I$ where negative (positive) $\Delta I$ favors the CW (CCW) mode.  The spectra in each panel are normalized to the maximum intensity point of that panel.
}
\label{spectra}
\end{figure}
The spectral properties of the racetrack laser and the impact that pumping of the waveguide arms has on them are summarized in Fig. \ref{spectra}. We measure the signal exiting one of the waveguide arms using a Fourier Transform Infrared Spectrometer (FTIR) with \SI{0.125}{{\cm}^{-1}} spectral resolution. The pure racetrack spectrum with no current applied to either waveguide arm is shown in the left panel and it reveals a series of longitudinal modes spanning $\approx$ \SI{14}{{\cm}^{-1}} with a measured free-spectral range (FSR) of $\approx$ \SI{0.70}{\cm^{-1}} which is an excellent match for the calculated value $1/nL$ = \SI{0.701}{\cm^{-1}}, where $L = \SI{4.318}{\mm}$ is the designed racetrack cavity length and $n = 3.3$ is the effective refractive index of the ridge waveguide. On the right panel of Fig.~\ref{spectra} we show the CW spectra as a function of current imbalance between the arms $\Delta I$ as well as average current $I_0$ which is marked at the top. In this case, a positive $\Delta I$ means higher current is applied to the waveguide arm favoring CCW operation which explains why the spectra within each panel are generally weaker for $\Delta I > 0$.

The FSR, once some current is applied to the waveguide arms, remains the same as for the pure racetrack spectrum, although there is a small red shift in the center frequency which can be explained by increased local temperature of the chip due to the applied current $2I_0$. For low values of $I_0$, despite the red shift, the spectral envelope that we measure remains similar to the racetrack spectrum without injected current. As $I_0$ increases this envelope changes significantly with only a few modes becoming dominant as can be seen for $I_0 = $ \SI{790}{\mA} and $I_0 = $ \SI{900}{\mA}. For these high values of $I_0$ we are able to recover the pure racetrack spectrum around $\Delta I = 0$ mark. This is most visible for $I_0 =$ \SI{790}{\mA} where even though only a couple of modes dominate for negative $\Delta I$, around the 0 mark, we recover the original envelope. Hence, lower values of current applied to the waveguide arms do not alter the spectral envelope of the racetrack laser; however, for higher values the envelope changes significantly. When the current differential between the arms is relatively small, the resulting spectra are also similar to the pure racetrack spectrum envelope. 
%The calculated FSR for the entire waveguide section is \SI{0.367}{cm^{-1}}. 

\section{Conclusion}
In this paper we have shown the operation of a novel monolithic QCL-based system comprising of a racetrack laser and an evanescently-coupled bus waveguide with independently-controlled arms based on a specially-designed tapered coupler. We demonstrate deterministic and controllable selection between the two counter-propagating racetrack modes via spontaneous emission coupling from the differentially pumped waveguide arms. As little as \SI{14}{\mA} of current imbalance, which represents $\approx 1.6\%$ of the average input current into the waveguides, can trigger a full and permanenet switch between the CW and CCW modes. By further investigating the mechanism of mode selection, we have demonstrated the possibility of bistable operation of our device. Because of the bistable nature of the ring, the waveguide arms do not need constant pumping in order to maintain a particular directional mode. In fact, a small perturbation is sufficient to select the mode and the racetrack laser will maintain this mode even after the input perturbation dissipates. 

Finally, we have demonstrated an efficient outcoupling mechanism by measuring up to $\approx$ \SI{4}{\mA} of optical power coupled out of the racetrack laser. These results bring us closer towards practical applications of controllable unidirectional ring or racetrack QCLs. 

% \section{Backmatter}

%\begin{backmatter}
\paragraph{Funding} This work was supported by the Eric and Wendy Schmidt Transformative Technology Fund. S.K. acknowledges support from the Yan Huo *94 Graduate Fellowship.

\paragraph{Acknowledgments} We acknowledge the Micro and Nano Fabrication Center (MNFC) of Princeton University where fabrication was carried out.

\paragraph{Disclosures} The authors declare no conflicts of interest.

\end{document}